\documentclass{article}
\usepackage{graphicx}
\usepackage[centertags]{amsmath}
\usepackage{amsfonts}
\usepackage{amssymb}
\usepackage{amsthm}
\newcommand{\rd}{\mathrm{d}}

\title{Mechanics of Incompressible Test Bodies\\
Moving in Riemannian Spaces}

\author{Vasyl~Kovalchuk, Barbara~Go\l ubowska, Ewa~Eliza~Ro\.{z}ko\\
Institute of Fundamental Technological Research\\
Polish Academy of Sciences\\
$5^{\rm B}$, Pawi\'{n}skiego str., 02-106 Warsaw, Poland\\
e-mails: vkoval@ippt.pan.pl, bgolub@ippt.pan.pl,\\
erozko@ippt.pan.pl}

\begin{document}

\maketitle

\begin{abstract}
In the present paper we have discussed the mechanics of incompressible test bodies moving in Riemannian spaces with non-trivial curvature tensors. For Hamilton's equations of motion the solutions have been obtained in the parametrical form and the special case of the purely gyroscopic motion on the sphere has been discussed. For the geodetic case when the potential is equal to zero the comparison between the geodetic and geodesic solutions has been done and illustrated in details for the case of a particular choice of the constants of motion of the problem.

The obtained results could be applied, among others, in geophysical problems, e.g., for description of the movement of continental plates or the motion of a drop of fat or a spot of oil on the surface of the ocean (e.g., produced during some ``ecological disaster''), or generally in biomechanical problems, e.g., for description of the motion of objects with internal structure on different curved two-dimensional surfaces (e.g., transport of proteins along the curved biological membranes).
\end{abstract}

\section*{Introduction}

In our previous papers \cite{Gol_01,Gol_04,SG_10,all_06} we have discussed in details different aspects of the mechanics of infinitesimal test bodies moving in Riemannian spaces with non-trivial curvature tensors. It is generally well known that for a given $n$-dimensional Riemannian space its isometry group has dimension $k\leq n(n+1)/2$, whereas the maximal possible dimension is attained for spaces of constant curvature \cite{KN_63,MF_80,Wolf_72}. That is mainly the reason why we have chosen as an instructive example in the present paper the most symmetrical case of the spherical surface which has simultaneously constant positive Gaussian and mean curvatures.

Let us also mention that there is plenty of microscopic models of continua that are dealing with material points with attached to them geometric objects (collective or internal degrees of freedom), e.g., liquid crystals that can be described as continua of infinitesimal rods \cite{Cap_89,CM_03,Mar_00}. Therefore our infinitesimal test bodies are quite legitimate objects that can appear in mathematical modelling of many real-world situations. For instance, in description of the motion of some pollution regions (drops of fat or spots of oil spilled from damaged tankers during some ``ecological disasters'') on the oceanic surface (modelled as relatively small two-dimensional bodies on the spherical surface) or in description of the motion of objects with internal structure on different curved two-dimensional surfaces (e.g., transport of proteins along the curved biological membranes).

This work is a continuation of our recent research where we have investigated the motion of infinitesimal gyroscopes (rigid bodies) on such very interesting and instructive two-dimensional surfaces as Delaunay surfaces (spheres and cylinders as limiting cases of unduloids) of constant mean curvature \cite{KGM_19} or Mylar balloons \cite{KM_20}. In the present article we are generalizing the description to the situation of the incompressible test bodies for which apart from rotations also some deformation is allowed. Let us also describe the subject of our interest in the two-fold manner, i.e., from the very beginning let us introduce some general formulation (independent of the particular form of the two-dimensional surface on which the test body is moving) and simultaneously illistrate the general procedure on the example of an incompressible test body moving on a two-dimensional spherical surface embedded into some three-dimensional Euclidean space.

Let us start with the introduction of the tripple $(M,\Gamma,g)$, where $M$ is denoting a differential manifold (e.g., a two-dimensional sphere) endowed with some affine connection $\Gamma$ and metric tensor $g$ (they can be interrelated or not). For an infinitesimal affinely-rigid (homogeneously deformable) test body we have that $x\in M$ represents the spatial position of the body ``as a whole'' (it is a remnant of the centre of mass position in the flat-space theory), whereas the internal configuration (additional variables attached at the spatial position of the body) of such a homogeneously deformable body is injected into the tangent space $T_{x}M$ (microphysical space) where it can be identified with linear frames (ordered bases) $e_{A}\in T_{x}M$. In this way the configuration space of the infinitesimal homogeneously deformable test body moving in the physical space $M$ is given by the manifold $FM$ of linear frames in $M$
\begin{equation}\label{eq01}
Q=FM=\bigcup_{x\in M}F_{x}M,
\end{equation}
where $F_{x}M$ denotes the manifold of linear frames in the tangent space $T_{x}M$. Then any system of local coordinates $x^{i}$ intorduced on $M$ induces local coordinates  $(x^{i},e^{i}{}_{A})$ on $FM$ and local coordinates $(x^{i},e^{A}{}_{i})$ on $F^{\ast}M$, which is the manifold of all linear co-frames in $M$.

From the mechanical point of view, we have that the action of the structural group GL$(n,\mathbb{R})$ for our infinitesimal homogeneously deformable test body on $Q=FM$ and $Q^{\ast}=F^{\ast}M$ corresponds to ``micromaterial'' transformations that can be seen as the infinitesimal limit of the usual material transformations. This means that $e\in F_{x}M$ is canonically identical with some linear isomorphism of $\mathbb{R}^{n}$ onto $T_{x}M$, whereas $\widetilde{e}\in F_{x}^{\ast}M$ is canonically identical with a linear isomorphism of $T_{x}M$ onto $\mathbb{R}^{n}$. In this way $\mathbb{R}^{n}$ (additionally equipped with the metric tensor $\eta$) plays the role of the ``micromaterial'' space (corresponding Lagrange variables) and $T_{x}M$ plays the role of the ``microphysical'' space (corresponding Euler variables).

In the general case of unconstrained infinitesimal homogeneously deformable test bodies $\left(x^{i},e^{i}{}_{A}\right)$ are ``good'' independent (unconstrained) generalized coordinates. However, after some constraints are imposed, the quantities $e^{i}{}_{A}$ are no longer independent and cannot be used as generalized coordinates. For example, in the case of incompressibility constraints we have that \cite{Slaw_82,all_06}
\begin{equation}\label{eq02}
\det\left[e^{i}{}_{A}\right]=\sqrt{\frac{\det\left[g_{ij}\right]}{\det\left[\eta_{AB}\right]}}.
\end{equation}
When the metric tensors $g$ and $\eta$ are both given with the help of identity matrices (the flat-space situation), then in the right-hand side of (\ref{eq02}) we have simply $1$. In our situation the right-hand side can be simplified to $\sqrt{\det\left[g_{ij}\right]}$ because the micromaterial space is chosen to be some Euclidean space $\mathbb{R}^{n}$ with $\eta_{AB}=\delta_{AB}$.

Nevertheless, some geometric techniques based on the use of orthonormal aholonomic reference frames may be developed. In this situation we can choose on $M$ some pre-established, fixed once and for all fields of linear orthonormal aholonomic frames $E$ for which we have the following relation
\begin{equation}\label{eq03}
g\left(E_{A},E_{B}\right)=g_{ij}E^{i}{}_{A}E^{j}{}_{B}=\eta_{AB}.
\end{equation}
The dual co-frames $\widetilde{E}=(\ldots,E^{A},\ldots)$ are orthonormal with respect to the inverse (contravariant) metrics $\widetilde{g}$ and $\widetilde{\eta}$ defined on the differential manifold $M$ and micromaterial space $\mathbb{R}^{n}$ respectively, i.e., 
\begin{equation}\label{eq04}
\widetilde{g}\left(E^{A},E^{B}\right)=E^{A}{}_{i}E^{B}{}_{j}g^{ij}=\eta^{AB}.
\end{equation}
Their particular choice, both technically convenient and geometrically lucid, in most cases is dictated by the structure of a given Riemannian space $(M,g)$. So, in this way we obtain that in the very description of the considered problem is to some extent incorporated the information about the intrinsic geometry of the differential manifold $M$ (defined by its first fundamental form in the situation when it is embedded into the higher-dimensional space), including the information about its curvature.

For example, for the considered situation of the two-dimensional sphere embedded into the three-dimensional Euclidean space, i.e., $S^{2}(0,R)\subset\mathbb{R}^{3}$, we can use the following parameterization (a particular choice of local coordinates $\mathbf{X}=\left(x,y,z\right)$ in the physical space $M$):
\begin{equation}\label{eq05}
x=R\sin u\cos v,\quad y=R\sin u\sin v,\quad z=R\cos u,\quad  x^{2}+y^{2}+z^{2}=R^{2},
\end{equation}
where $R$ is the fixed radius of the sphere and $u\in[0,\pi]$, $v\in[0,2\pi)$ are the Euler angles in $\mathbb{R}^{3}$ (when $u=0$ corresponds to the ``North Pole'' and $u=\pi$ corresponds to the ``South Pole'' of our sphere and the condition $v={\rm const}$ defines the meridians). For the parameterization (\ref{eq05}) we obtain that the first fundamental form ${\rm I}=\left\{E,F,G\right\}$ has the following components:
\begin{eqnarray}
&&E=g_{uu}=\left(\mathbf{X}_{u},\mathbf{X}_{u}\right)=R^2,\label{eq06}\\
&&F=g_{uv}=\left(\mathbf{X}_{u},\mathbf{X}_{v}\right)=g_{vu}=\left(\mathbf{X}_{v},\mathbf{X}_{u}\right)=0,\label{eq07}\\ 
&&G=g_{vv}=\left(\mathbf{X}_{v},\mathbf{X}_{v}\right)=R^2\sin^2 u,\label{eq08}
\end{eqnarray}
where $\mathbf{X}_{i}$ denotes the derivative of $\mathbf{X}$ with respect to the local coordinate $x^i$.

From the other side, the orthonormal aholonomic reference frames $E_{A}$ can be chosen in the following way (dependent on the above-defined particular choice of local coordinates in the physical space $M$):
\begin{eqnarray}
&&E^{u}{}_{u}=\frac{1}{\sqrt{g_{uu}}}\frac{\partial}{\partial u}=\frac{1}{R}\frac{\partial}{\partial u},\quad E^{v}{}_{u}=E^{u}{}_{v}=0,\label{eq09}\\
&&E^{v}{}_{v}=\frac{1}{\sqrt{g_{vv}}}\frac{\partial}{\partial v}=\frac{1}{R\sin u}\frac{\partial}{\partial v}.\label{eq10}
\end{eqnarray}

So, dynamically speaking, at any time instant $t\in \mathbb{R}$ the body is instantaneously placed at the geometric point $\mathbf{X}(t)\in M$ and its internal configuration is described by linear frames $e(t)_{A}=E\left[\mathbf{X}(t)\right]_{B}\varphi(t)^{B}{}_{A}$. In this way, instead of describing the motion in terms of time-dependent quantities $\left(x(t)^{i},e(t)^{i}{}_{A}\right)$, we can describe it in terms of quantities $\left(x(t)^{i},\varphi(t)^{A}{}_{B}\right)$. 

\section{Affine velocities for internal motion}

In our previous papers on the afiine bodies \cite{Slaw_82,all_06,all_11,all_12} we have shown that apart from the quantities $\varphi(t)$ it is also convinient to introduce the so-called affine velocities $\Omega$ that are corresponding to the internal motion of our infinitesimal homogeneously deformable test body. They are defined in the co-moving representation by the following equation:
\begin{equation}\label{eq11}
\frac{De_{A}}{Dt}=e_{B}\widehat{\Omega}^{B}{}_{A}.
\end{equation}
In order to write them explicitly, we must first find formulas for the covariant derivatives in terms of the quantities $E$ and $\varphi$, i.e.,
\begin{equation}\label{eq12}
\frac{De_{A}}{Dt}=\frac{D}{Dt}\left(E_{B}\varphi^{B}{}_{A}\right)=\frac{DE_{B}}{Dt}\varphi^{B}{}_{A}+E_{B}\frac{\rd \varphi^{B}{}_{A}}{\rd t}\cdot
\end{equation}
Let us now express the along-curve differentiation of $E_{B}$ through its field-differentiation (which is well defined because $E$ is given as a global vector field on $M$), i.e.,
\begin{equation}\label{eq13}
\frac{DE_{B}}{Dt}=\left(\nabla_{i}E_{B}\right) \frac{\rd x^{i}}{\rd t}=\left(\nabla_{E_{C}}E_{B}\right)E^{C}{}_{i}\frac{\rd x^{i}}{\rd t}\cdot
\end{equation}
It is more convenient to use auxiliary aholonomic coefficients of our affine connection $\Gamma$ with respect to the fields $E$, i.e., 
\begin{equation}\label{eq14}
\nabla_{E_{C}}E_{B}=\Gamma^{A}{}_{BC}E_{A}. 
\end{equation}

In this way the usual holonomic coefficients of $\Gamma$ with respect to coordinates $x^{i}$ are given as
\begin{equation}\label{eq15}
\Gamma^{i}{}_{jk}=E^{i}{}_{A}\Gamma^{A}{}_{BC}E^{B}{}_{j}E^{C}{}_{k}
+E^{i}{}_{A}E^{A}{}_{j,k}=E^{i}{}_{A}\Gamma^{A}{}_{BC}E^{B}{}_{j}E^{C}{}_{k}+\Gamma[E]^{i}{}_{jk},
\end{equation}
where the second term denotes the teleparallelism connection induced by the fields $E$, i.e., $\nabla_{[\Gamma[E]]}E_{A}=0$. In other words, it is the only affine connection with respect to which all the fields $E_{A}$ and their dual co-fields $E^{A}$ are parallel, whereas its curvature tensor vanishes and the corresponding torsion tensor is
\begin{equation}\label{eq16}
S[E]^{i}{}_{jk}=E^{i}{}_{A}E^{A}{}_{[j,k]}=\frac{1}{2}E^{i}{}_{A}\left(E^{A}{}_{j,k}-E^{A}{}_{k,j}\right).
\end{equation}

For the sphere's case, when the metric tensor's components $g_{ij}$ are defined by formulas (\ref{eq06})--(\ref{eq08}), we can calculate explicitly the Levi-Civita affine connection's coefficients $\Gamma^{i}{}_{jk}$ with respect to coordinates $x^{i}$ given by (\ref{eq05}) as follows:
\begin{equation}\label{eq17}
\Gamma^{i}{}_{jk}=\left\{\begin{array}{c} i \\ jk
\end{array}\right\}=
\frac{1}{2}g^{im}\left(g_{mj,k}+g_{mk,j}-g_{jk,m}\right)
\end{equation}
with only three of the eight components being non-zero, i.e.,
\begin{eqnarray} \label{eq18}
\Gamma^{u}{}_{vv}=-\sin u\cos u,\qquad \Gamma^{v}{}_{uv}=\Gamma^{v}{}_{vu}=\cot u.
\end{eqnarray}

Next for the reference frames $E_{A}$ defined by formulas (\ref{eq09})--(\ref{eq10}) the aholonomic coefficients $\Gamma^{A}{}_{BC}$ from (\ref{eq14}) can be explicitly calculated as follows:
\begin{eqnarray} \label{eq19}
\Gamma^{A}{}_{BC}=E^{A}{}_{i}\left(\Gamma-\Gamma[E]\right)^{i}{}_{jk}E^{j}{}_{B}E^{k}{}_{C}
\end{eqnarray}
with only two of the eight components being non-zero, i.e.,
\begin{eqnarray} \label{eq20}
-\Gamma^{u}{}_{vv}=\Gamma^{v}{}_{uv}=\frac{1}{R}\cot u.
\end{eqnarray}

Finally, inserting the expression (\ref{eq13})--(\ref{eq15}) and (\ref{eq19})--(\ref{eq20}) into the expression (\ref{eq12}) we obtain that the co-moving affine velocity $\widehat{\Omega}$ may be given as the sum of the ``drift'' (or ``drive'') term $\widehat{\Omega}_{\rm dr}$ which describes the time rate of the part of internal motion that is contained in the very fields $E_{A}$ themselves, i.e.,
\begin{equation}\label{eq21}
\widehat{\Omega}_{\rm dr}{}^{A}{}_{B}=\varphi^{-1}{}^{A}{}_{C}\widehat{\chi}_{\rm dr}{}^{C}{}_{D}\varphi^{D}{}_{B},
\end{equation}
where the matrix $\widehat{\chi}_{\rm dr}$ can be defined with the help of the two-dimensional non-singular skew-symmetric matrix $S=\left[\begin{array}{cc}
0 & 1\\
-1 & 0
\end{array}\right]$ for which $\det S=1$ as follows:
\begin{equation}\label{eq22}
\widehat{\chi}_{\rm dr}{}^{C}{}_{D}=\Gamma^{C}{}_{DF}E^{F}{}_{i}\frac{\rd x^{i}}{\rd t}=\cos u\ \frac{\rd v}{\rd t}\ S^{C}{}_{D},
\end{equation}
and the ``relative'' term $\widehat{\Omega}_{\rm rl}$ which refers to the part of internal motion performed with respect to the just passed prescribed reference frames $E_{A}$, i.e.,
\begin{equation}\label{eq23}
\widehat{\Omega}_{\rm rl}{}^{A}{}_{B}=\varphi^{-1}{}^{A}{}_{C}\frac{\rd \varphi^{C}{}_{B}}{\rd t}\cdot
\end{equation}
The shifting of indices in $\widehat{\Omega}^{A}{}_{B}=\widehat{\Omega}_{\rm dr}{}^{A}{}_{B}+\widehat{\Omega}_{\rm rl}{}^{A}{}_{B}$ can be done with the help of the metric tensor $\eta_{AB}$. We can also introduce its spatial representation, i.e., $\Omega^{i}{}_{j}=e^{i}{}_{A}\widehat{\Omega}^{A}{}_{B}e^{B}{}_{j}$, in which we can shift indices using the metric tensor $g_{ij}$. 

\section{Two-polar decomposition of configurations}

In some situations (e.g., when the body is isotropic in the micromaterial space), it is convenient to use the two-polar decomposition of the configuration matrix \cite{all_06}, i.e., for the incompressibility constraints we have that $\varphi=LDR^{-1}$, where $L$ and $R$ are orthogonal matrices and $D$ is a diagonal one of the following form:
\begin{equation}\label{eq24}
L=\left[\begin{array}{cc}
\cos \alpha & -\sin\alpha\\
\sin\alpha & \cos\alpha
\end{array}\right],\ D=\left[\begin{array}{cc}
e^{\lambda} & 0\\
0 & e^{-\lambda}
\end{array}\right],\ R=\left[\begin{array}{cc}
\cos\beta & -\sin\beta\\
\sin\beta & \cos\beta
\end{array}\right].
\end{equation}
Of course, the incompressibility constraints may be described analytically as the requirement that the affine velocity should be traceless, i.e.,
\begin{equation}\label{eq25}
{\rm Tr}\ \widehat{\Omega}=\widehat{\Omega}^{A}{}_{A}=\widehat{\Omega}_{\rm dr}{}^{A}{}_{A}+\widehat{\Omega}_{\rm rl}{}^{A}{}_{A}=0,
\end{equation}
and really we can easily check that in the two-polar decomposition (\ref{eq24}) we have
\begin{eqnarray}
&&\widehat{\Omega}_{\rm dr}{}^{A}{}_{A}=\varphi^{-1 A}{}_{C}\widehat{\chi}_{\rm dr}{}^{C}{}_{D}\varphi^{D}{}_{A}=\widehat{\chi}_{\rm dr}{}^{C}{}_{C}=\cos u\ \dot{v}\ S^{C}{}_{C}=0,\label{eq26}\\
&&\widehat{\Omega}_{\rm rl}{}^{A}{}_{A}=\varphi^{-1}{}^{A}{}_{C}\dot{\varphi}^{C}{}_{A}=\widehat{\chi}_{\rm rl}{}^{A}{}_{A}+D^{-1}{}^{A}{}_{C}\dot{D}^{C}{}_{A}-\widehat{\vartheta}_{\rm rl}{}^{A}{}_{A}=0\label{eq27}
\end{eqnarray}
because the co-moving angular velocities of the left and right fictitious gyroscopes in the two-polar decomposition (\ref{eq24})
\begin{equation}\label{eq28}
\widehat{\chi}_{\rm rl}=L^{-1}\dot{L}=\dot{\alpha}S^{T},\qquad 
\widehat{\vartheta}_{\rm rl}=R^{-1}\dot{R}=\dot{\beta}S^{T}
\end{equation}
are given by the skew-symmetric (therefore, also traceless) matrices $S^{T}$, whereas
\begin{equation}\label{eq29}
D^{-1}\dot{D}=\left[\begin{array}{cc}
e^{-\lambda} & 0\\
0 & e^{\lambda}
\end{array}\right]\left[\begin{array}{cc}
e^{\lambda} & 0\\
0 & -e^{-\lambda}
\end{array}\right]\dot{\lambda}=\left[\begin{array}{cc}
1 & 0\\
0 & -1
\end{array}\right]\dot{\lambda}
\end{equation}
is given by the traceless matrix as well.

Explicitly in the two-polar decomposition (\ref{eq24}) we have that the components of the total affine velocity $\widehat{\Omega}$ are given by the following expression:
\begin{equation}\label{eq30}
\widehat{\Omega}=\widehat{\Omega}_{\rm dr}+\widehat{\Omega}_{\rm rl}=R\left(D^{-1}\widehat{\chi}D+D^{-1}\dot{D}-\widehat{\vartheta}_{\rm rl}\right)R^{-1},
\end{equation}
where for the sphere's case (\ref{eq05}) the expression $\widehat{\chi}=\widehat{\chi}_{\rm dr}+\widehat{\chi}_{\rm rl}=\left(\dot{\alpha}-\cos u\ \dot{v}\right)S^{T}$ contains the drift (\ref{eq22}) and relative (\ref{eq28}) terms of the $L$-rotation, while the $R$-rotation has no drift term $\widehat{\vartheta}_{\rm dr}=0$ and only relative one $\widehat{\vartheta}_{\rm rl}$. In this way, we see that the left fictitious gyroscope $L$ alone absorbs the whole information about the geometry of the physical space $M$ and leaves the right fictitious gyroscope $R$ to be geometry-independent.

Let us note that from the incompressibility constraints we can easily obtain the special case of the purely gyroscopic motion when in the two-polar decomposition (\ref{eq24}) we suppose that $\lambda=\beta=0$. In this case the co-moving affine velocity $\widehat{\Omega}$ becomes an angular velocity $\widehat{\omega}$ (which is given by the skew-symmetric matrix $S$) and for the sphere's case (\ref{eq05}) it is given as follows:
\begin{equation}\label{eq31}
\widehat{\omega}=\widehat{\omega}_{\rm dr}+\widehat{\omega}_{\rm rl}=\left(\dot{\alpha}-\cos u\ \dot{v}\right)S^{T}.
\end{equation}

\section{D'Alembert models of kinetic energy}

Let us now consider the traditional d'Alembert method of deriving the kinetic energy \cite{Slaw_82,all_06} for which the Lagrangian of our infinitesimal homogeneously deformable test body moving in a Riemannian space (e.g., sphere) can be given as follows:
\begin{equation}\label{eq32}
L=\frac{m}{2}g_{ij}\frac{\rd x^i}{\rd t}\frac{\rd x^j}{\rd t}+\frac{1}{2}g_{ij}\frac{De^{i}{}_{A}}{Dt}\frac{De^{j}{}_{B}}{Dt}J^{AB}-V\left(u,\lambda\right),
\end{equation}
where $m$ denotes the mass of the infinitesimal test body, $J\in\mathbb{R}^{n}\otimes\mathbb{R}^{n}$ is the symmetric and positively definite micromaterial inertial tensor (postulated as something primary) describing the internal properties of our infinitesimal test body, whereas $V$ is some potential term well-suited to the geometry of the physical space $M$ (e.g., sphere) and it is dependent only on the translational degree of freedom $u$ and the deformation invariant $\lambda$ from the two-polar decomposition (\ref{eq24}) (therefore, the variables $v$, $\alpha$, and $\beta$ are cyclic coordinates, i.e., they do not occur explicitly in the corresponding equations of motion).

In the above expression the total kinetic energy $T$ is postulated as the sum of the translational part which for the sphere's case (\ref{eq05}) has the following form:
\begin{equation}\label{eq33}
T_{{\rm tr}}=\frac{mR^2}{2}\left[\dot{u}^{2}+\sin^{2} u\ \dot{v}^{2}\right],
\end{equation}
and the internal part which using (\ref{eq11}) can be rewritten as follows:
\begin{equation}\label{eq34}
T_{{\rm int}}=\frac{1}{2}G[e]_{AB}\widehat{\Omega}^{A}{}_{C}\widehat{\Omega}^{B}{}_{D}J^{CD},
\end{equation}
where $G[e]_{AB}=g_{ij}e^{i}{}_{A}e^{j}{}_{B}=g_{ij}E^{i}{}_{C}E^{j}{}_{D}\varphi^{C}{}_{A}\varphi^{D}{}_{B}=\eta_{CD}\varphi^{C}{}_{A}\varphi^{D}{}_{B}$ is the Green deformation tensor defined in the micromaterial space $\mathbb{R}^{n}$. 

In the two-polar decomposition (\ref{eq24}) we would have that
\begin{equation}\label{eq35}
G[e]=\varphi^{T}\varphi=RD^{2}R^{-1}={\rm Id}_{2}\cosh(2\lambda)+\Lambda(2\beta)\sinh(2\lambda),
\end{equation}
where ${\rm Id}_{2}$ is the identity matrix and $\Lambda(\cdot)=\left[\begin{array}{cc}
\cos(\cdot) & \sin(\cdot)\\
\sin(\cdot) & -\cos(\cdot)
\end{array}\right]$ is traceless.

Therefore, using (\ref{eq30}), (\ref{eq35}) and supposing that our infinitesimal test body is isotropic in the micromaterial space, i.e., in two dimensions we have that $J^{AB}=\left(I/2\right)\eta^{AB}$, therefore, ${\rm Tr}\left(\eta J\right)=(I/2){\rm Tr}\left({\rm Id}_2\right)=I$, we obtain that the expression for the internal kinetic energy in the sphere's case (\ref{eq05}) is as follows:
\begin{eqnarray}\label{eq36}
T_{\rm int}&=&\frac{I}{4}{\rm Tr}\left(\widehat{\Omega}^{T}G[e]\widehat{\Omega}\right)=
\frac{I}{4}{\rm Tr}\left[\left(\dot{D}-D\widehat{\chi}+\widehat{\vartheta}_{\rm rl}D\right)\left(\dot{D}+\widehat{\chi}D-D\widehat{\vartheta}_{\rm rl}\right)\right]\nonumber\\
&=&\frac{I}{4}{\rm Tr}\left(\dot{D}^2\right)-\frac{I}{4}{\rm Tr}\left(D^2\left[\widehat{\chi}^{2}+\widehat{\vartheta}^{2}_{\rm rl}\right]\right)+\frac{I}{2}{\rm Tr}\left(D\widehat{\chi}D\widehat{\vartheta}_{\rm rl}\right)\nonumber\\
&=&\frac{I}{2}\cosh(2\lambda)\left[\dot{\lambda}^2+\left(\dot{\alpha}-\cos u\ \dot{v}\right)^2+\dot{\beta}^{2}\right]-I\left(\dot{\alpha}-\cos u\ \dot{v}\right)\dot{\beta}.
\end{eqnarray}

For the special case of the purely gyroscopic motion (\ref{eq31}) we obtain that the Green deformation tensor is equal to the metric tensor in the micromaterial space, i.e., $G[e]=\eta$, the internal kinetic energy (\ref{eq36}) is reduced to the expression $(I/2)\left(\dot{\alpha}-\cos u\ \dot{v}\right)^2$, therefore, the Lagrangian (\ref{eq32}) can be written as follows:
\begin{equation}\label{eq37}
L=\frac{mR^2}{2}\left[\dot{u}^{2}+\sin^{2} u\ \dot{v}^{2}\right]+\frac{I}{2}\left(\dot{\alpha}-\cos u\ \dot{v}\right)^2-V\left(u\right).
\end{equation}

Equivalently, we can rewrite the total kinetic energy in the following form: 
\begin{equation}\label{eq38}
T=T_{\rm tr}+T_{\rm int}=\frac{m}{2}G(q)_{ij}\frac{dq^{i}}{dt}\frac{dq^{j}}{dt},
\end{equation}
where the generalized coordinates are ordered as $\left\{q^{i}\right\}=\left\{u,v,\alpha,\beta,\lambda\right\}$ and the metric-like matrix $G(q)_{ij}$ for the sphere's case (\ref{eq05}) is given as follows:
\begin{equation}\label{eq39}
G(q)_{ij}=\left[\begin{array}{ccc}
R^2 & 0_{3}  & 0\\
0_{3} & M_{3} & 0_{3}\\
0 & 0_{3} & \frac{I}{m}\cosh(2\lambda)
\end{array}\right]
\end{equation}
with $0_{3}$ being the three-dimensional zero matrix and the matrix $M_{3}$ is given as
\begin{equation}\label{eq40}
\left[\begin{array}{ccc}
R^2\sin^2 u+\frac{I}{m}\cosh(2\lambda)\cos^2 u & -\frac{I}{m}\cosh(2\lambda)\cos u & \frac{I}{m}\cos u \\
-\frac{I}{m}\cosh(2\lambda)\cos u & \frac{I}{m}\cosh(2\lambda) & -\frac{I}{m}\\
\frac{I}{m}\cos u & -\frac{I}{m} & \frac{I}{m}\cosh(2\lambda)
\end{array}\right].
\end{equation}
The inverse matrix to (\ref{eq39}) has the following form:
\begin{equation}\label{eq41}
G(q)^{ij}=\left[\begin{array}{ccc}
\frac{1}{R^2} & 0_{3}  & 0\\
0_{3} & M^{-1}_{3} & 0_{3}\\
0 & 0_{3} & \frac{m}{I\cosh(2\lambda)}
\end{array}\right]
\end{equation}
with the matrix $M^{-1}_{3}$ is given as
\begin{equation}\label{eq42}
\left[\begin{array}{ccc}
\frac{1}{ R^2\sin^2 u} & \frac{\cos u}{R^2\sin^2 u} & 0 \\
\frac{\cos u}{R^2\sin^2 u} & \frac{\cos^2 u}{R^2\sin^2 u}+\frac{m}{I}\frac{\cosh(2\lambda)}{\sinh^2(2\lambda)} & \frac{m}{I}\frac{1}{\sinh^2(2\lambda)}\\
0 & \frac{m}{I}\frac{1}{\sinh^2(2\lambda)} & \frac{m}{I}\frac{\cosh(2\lambda)}{\sinh^2(2\lambda)}
\end{array}\right].
\end{equation}
From the expressions (\ref{eq39}) and (\ref{eq41}) we can easily obtain the form of the Legendre transformation $p_{i}=\partial L/\partial \dot{q}^i=mG(q)_{ij}\dot{q}^j$, where
\begin{eqnarray}
&&p_{u}=mR^2\dot{u}, \qquad p_{\lambda}=I\cosh(2\lambda)\dot{\lambda},\label{eq43}\\
&&p_{v}=\left[mR^2\sin^2 u+I\cosh(2\lambda)\cos^2 u\right]\dot{v}-I\cosh(2\lambda)\cos u\dot{\alpha}+I\cos u\dot{\beta},\qquad\label{eq44}\\
&&p_{\alpha}=-I\cosh(2\lambda)\cos u\dot{v}+I\cosh(2\lambda)\dot{\alpha}-I\dot{\beta},\label{eq45}\\
&&p_{\beta}=I\cos u\dot{v}-I\dot{\alpha}+I\cosh(2\lambda)\dot{\beta},\label{eq46}
\end{eqnarray}
and the inverse Legendre transformation $\dot{q}^i=(1/m)G(q)^{ij}p_{j}$, where
\begin{eqnarray}
&&\dot{u}=\frac{p_{u}}{mR^2},\qquad \dot{\alpha}=\frac{\cos u\left(p_{v}+\cos u\ p_{\alpha}\right)}{mR^2 \sin^2 u}+\frac{\cosh(2\lambda)p_{\alpha}+\ p_{\beta}}{I\sinh^2(2\lambda)},\label{eq47}\\
&&\dot{v}=\frac{p_{v}+\cos u\ p_{\alpha}}{mR^2 \sin^2 u},\quad \dot{\beta}=\frac{p_{\alpha}+\cosh(2\lambda)\ p_{\beta}}{I\sinh^2(2\lambda)},\quad \dot{\lambda}=\frac{p_{\lambda}}{I\cosh(2\lambda)}.\qquad\label{eq48}
\end{eqnarray}
Finally, denoting that 
\begin{equation}\label{eq49}
\dot{\alpha}-\cos u\ \dot{v}=\frac{\cosh(2\lambda)p_{\alpha}+p_{\beta}}{I\sinh^2(2\lambda)}
\end{equation}
and substituting (\ref{eq47})--(\ref{eq49}) into the expression for the Lagrangian with the translational (\ref{eq33}) and internal (\ref{eq36})  kinetic energies we obtain the Hamiltonian $H(q,p)=\mathcal{T}(q,p)+V(u,\lambda)$ with the total kinetic energy given as follows:
\begin{eqnarray}\label{eq50}
\mathcal{T}(q,p)&=&\frac{\ p^{2}_{u}}{2mR^2}+\frac{\left(p_{v}+\cos u\ p_{\alpha}\right)^2}{2mR^2\sin^2 u}\nonumber\\
&+&\frac{p_{\lambda}^{2}}{2I\cosh(2\lambda)}+\frac{\left(p_{\alpha}+p_{\beta}\right)^2}{8I\sinh^{2}\lambda}+\frac{\left(p_{\alpha}-p_{\beta}\right)^2}{8I\cosh^{2}\lambda}.
\end{eqnarray}
Again, for the purely gyroscopic motion ($\lambda=\beta=0$) we have that
\begin{equation}\label{eq50a}
\mathcal{T}(q,p)=\frac{\ p^{2}_{u}}{2mR^2}+\frac{\left(p_{v}+\cos u\ p_{\alpha}\right)^2}{2mR^2\sin^2 u}+\frac{p^2_{\alpha}}{2I},
\end{equation}

\section{Some convenient models of potentials}

Let us consider the following convenient choices of the separable potentials $V\left(u,\lambda\right)=V_{u}\left(u\right)+V_{\lambda}\left(\lambda\right)$. We suggest the first term $V_{u}\left(u\right)$ to be well suited to the geometry of the considered problem, e.g., for the sphere's case we have that
\begin{equation}\label{eq51}
V_{u}(u)=f(u)\det\left[g^{ij}\right]=\frac{f(u)}{R^{4}\sin^{2} u},
\end{equation}
where $f(u)$ is some function of the variable $u$. Therefore, for the different choices of the function $f(u)$
\begin{equation}\label{eq52}
f_{1}(u)=\frac{R^{2}\varkappa_{1}}{2m},\qquad f_{2}(u)=\frac{R^{2}\varkappa_{2}\cos u}{m},\qquad 
f_{3}(u)=\frac{R^{2}\varkappa_{3}\cos^{2} u}{2m},
\end{equation}
where $\varkappa_{i}$ are some constants, we obtain a class of potentials $V_{u}(u)$ built of trigonometric functions, i.e.,
\begin{equation}\label{eq53}
V_{1}(u)=\frac{\varkappa_{1}}{2mR^{2}\sin^2 u},\quad V_{2}(u)=\frac{\varkappa_{2}\cos u}{mR^{2}\sin^2 u},\quad 
V_{3}(u)=\frac{\varkappa_{3}\cos^{2} u}{2mR^{2}\sin^2 u}.
\end{equation}
Similarly, we can also consider a certain class of potentials $V_{\lambda}\left(\lambda\right)$ built of hyperbolic functions ($\sigma_{i}$ are some constants), i.e.,
\begin{equation}\label{eq54}
V_{1}\left(\lambda\right)=\frac{\sigma_{1}}{2I\cosh(2\lambda)},\quad V_{2}(\lambda)=\frac{\sigma_{2}}{8I\sinh^2 \lambda},
\quad V_{3}(\lambda)=\frac{\sigma_{3}}{8I\cosh^2 \lambda}.
\end{equation}
Finally, for the both classes (\ref{eq53}) and (\ref{eq54}) we obtain the resulting effective shifting in the canonical momenta in the expression for the Hamiltonian function:
\begin{eqnarray}\label{eq55}
\mathcal{H}(q,p)&=&\frac{\ p^{2}_{u}}{2mR^2}+\frac{p^2_{v}+\varkappa_{1}+2\cos u\left(p_{v}p_{\alpha}+\varkappa_{2}\right)+\cos^2 u\left(p^{2}_{\alpha}+\varkappa_{3}\right)}{2mR^2\sin^2 u}\nonumber\\
&+&\frac{p_{\lambda}^{2}+\sigma_{1}}{2I\cosh(2\lambda)}+\frac{\left(p_{\alpha}+p_{\beta}\right)^2+\sigma_{2}}{8I\sinh^{2}\lambda}+\frac{\left(p_{\alpha}-p_{\beta}\right)^2+\sigma_{3}}{8I\cosh^{2}\lambda}.
\end{eqnarray}

\section{Hamilton's equations of motion}

In Hamiltonian mechanics the time evolution of the classical physical system is defined by the Hamilton's equations, i.e., for a given Hamiltonian function (\ref{eq55}) we obtain that the equations of motion can be calculated as follows:
\begin{equation}\label{eq56}
\dot{p}_{i}=-\frac{\partial \mathcal{H}}{\partial q^{i}},\qquad \dot{q}^{i}=\frac{\partial \mathcal{H}}{\partial p_{i}}.
\end{equation}
The second part of (\ref{eq56}) is essentially equivalent to the inverse Legendre transformation (\ref{eq47})--(\ref{eq48}) but the first part produces the following expressions:
\begin{eqnarray}
&&\dot{p}_{u}=-\frac{\partial}{\partial u}\left[\frac{p^2_{v}+\varkappa_{1}+2\cos u\left(p_{v}p_{\alpha}+\varkappa_{2}\right)+\cos^2 u\left(p^{2}_{\alpha}+\varkappa_{3}\right)}{2mR^2\sin^2 u}\right],\label{eq57}\\
&&\dot{p}_{\lambda}+\frac{p_{\lambda}^{2}}{2I}\frac{\partial}{\partial \lambda}\left[\frac{1}{\cosh(2\lambda)}\right]=-\frac{\partial}{\partial \lambda}\left[\frac{\sigma_{1}}{2I\cosh(2\lambda)}\right.\nonumber\\
&&\qquad\qquad\qquad\qquad\qquad\ +\left.\frac{\left(p_{\alpha}+p_{\beta}\right)^2+\sigma_{2}}{8I\sinh^{2}\lambda}+\frac{\left(p_{\alpha}-p_{\beta}\right)^2+\sigma_{3}}{8I\cosh^{2}\lambda}\right],\qquad\label{eq58}
\end{eqnarray}
whereas $\dot{p}_{v}=0$, $\dot{p}_{\alpha}=0$, $\dot{p}_{\beta}=0$, i.e., $p_{v}=mC_{2}$, $p_{\alpha}=IC_{3}$, $p_{\beta}=IC_{4}$ are constants of motion and $C_{i}$ are some integration constants.

Now the left-hand sides of (\ref{eq57}) and (\ref{eq58}) (understood as some functions of $u$ and $\lambda$) can be rewritten with the help of (\ref{eq47}) and (\ref{eq48}) as follows:
\begin{eqnarray}
&&\frac{p_{u}}{mR^2}\frac{\partial p_{u}}{\partial u}=\frac{\partial }{\partial u}\left[\frac{p^{2}_{u}}{2mR^2}\right],\label{eq59}\\
&&\frac{p_{\lambda}}{I\cosh(2\lambda)}\frac{\partial p_{\lambda}}{\partial \lambda}+\frac{p_{\lambda}^{2}}{2I}\frac{\partial}{\partial \lambda}\left[\frac{1}{\cosh(2\lambda)}\right]=\frac{\partial}{\partial \lambda}\left[\frac{p^{2}_{\lambda}}{2I\cosh(2\lambda)}\right].\label{eq60}
\end{eqnarray}
Combining the above expressions with the right-hand sides of (\ref{eq57}) and (\ref{eq58}) and performing the integration we obtain $p_{u}$ and $p_{\lambda}$ as the some functions of the variables $u$ and $\lambda$ respectively, i.e.,
\begin{eqnarray}
p^2_{u}(u)&=&C^{2}_{1}-\frac{p^2_{v}+\varkappa_{1}+2\cos u\left(p_{v}p_{\alpha}+\varkappa_{2}\right)+\cos^2 u\left(p^{2}_{\alpha}+\varkappa_{3}\right)}{\sin^2 u},\qquad\label{eq61}\\
\frac{p^2_{\lambda}(\lambda)}{\cosh(2\lambda)}&=&C^{2}_{5}-\frac{\sigma_{1}}{\cosh(2\lambda)}-\frac{\left(p_{\alpha}+p_{\beta}\right)^2+\sigma_{2}}{4\sinh^{2}\lambda}-\frac{\left(p_{\alpha}-p_{\beta}\right)^2+\sigma_{3}}{4\cosh^{2}\lambda},\label{eq62}
\end{eqnarray}
where $C_{1}$ and $C_{5}$ are integration constants. Substituting (\ref{eq61})-(\ref{eq62}) into (\ref{eq55}) we obtain that the energy of the considered system is also a constant of motion, i.e.,
\begin{equation}\label{eq63}
E=\frac{C^{2}_{1}}{2mR^2}+\frac{C^{2}_{5}}{2I}.
\end{equation}

Next let us express the first and last expressions in (\ref{eq47})--(\ref{eq48}) as follows:
\begin{equation}\label{eq64}
\frac{mR^2\rd u}{p_{u}(u)}=\frac{I\cosh(2\lambda)\rd \lambda}{p_{\lambda}(\lambda)}=\frac{\rd t}{1}.
\end{equation}
Substituting (\ref{eq61})--(\ref{eq62}) into the above expression (\ref{eq64}) we obtain that
\begin{eqnarray}
t(u)&=&mR^2\int\frac{\sin u\ \rd u}{\sqrt{A_{1}-2A_{2}\cos u-A_{3}\cos^2 u}},\label{eq65}\\
t(\lambda)&=&I\int\frac{\cosh(2\lambda)\sinh(2\lambda)\ \rd \lambda}{\sqrt{C^2_{5}\cosh^3(2\lambda)-B_{1}\cosh^2(2\lambda)-B_{2}\cosh(2\lambda)+\sigma_{1}}},\label{eq66}
\end{eqnarray}
where 
\begin{eqnarray}
&&A_{1}=C^{2}_{1}-p^{2}_{v}-\varkappa_{1},\quad A_{2}=p_{v}p_{\alpha}+\varkappa_{2},\quad A_{3}=C^{2}_{1}+p^{2}_{\alpha}+\varkappa_{3},\label{eq67}\\
&&B_{1}=p^{2}_{\alpha}+p^{2}_{\beta}+\sigma_{1}+\frac{\sigma_{2}+\sigma_{3}}{2},\quad B_{2}=C^{2}_{5}+2p_{\alpha}p_{\beta}+\frac{\sigma_{2}-\sigma_{3}}{2}.\qquad\label{eq68}
\end{eqnarray}
Changing the variables in the first integral to $x=-\cos u$ and in the second one to $y=\cosh(2 \lambda)$ we can rewrite (\ref{eq65})--(\ref{eq66}) as follows:
\begin{eqnarray}
t(x)&=&mR^2\int\frac{\rd x}{\sqrt{A_{1}+2A_{2}x-A_{3}x^2}},\label{eq69}\\
t(y)&=&\frac{I}{2}\int\frac{y\rd y}{\sqrt{C^2_{5}y^3-B_{1}y^2-B_{2}y+\sigma_{1}}}.\label{eq70}
\end{eqnarray}
In this way we parametrically represent the time variable $t$ through the translational variable $x$ (respectively $u$) or deformational one $y$ (respectively $\lambda$). Therefore, inverting the obtained functions $t(x)$, $t(y)$ after the proper integration of the expressions (\ref{eq69})--(\ref{eq70}) we finally obtain two solutions $x(t)$ and $y(t)$ (or equivallently $u(t)$ and $\lambda(t)$) of our equations of motion (\ref{eq56}). 

The time dependency of the other three angle variables $v$, $\alpha$, and $\beta$ can be obtain after substituting of (\ref{eq64}) into the corresponding expressions for their time derivatives from (\ref{eq47})--(\ref{eq48}) and integrating them with respect to $x$ and $y$:
\begin{eqnarray}
v(x)&=&\int\frac{p_{v}-p_{\alpha}x}{1-x^2}\frac{\rd x}{\sqrt{A_{1}+2A_{2}x-A_{3}x^2}},\label{eq71}\\
\alpha(x,y)&=&\int\frac{p_{\alpha}x-p_{v}}{1-x^2}\frac{x\rd x}{\sqrt{A_{1}+2A_{2}x-A_{3}x^2}}\nonumber\\
&+&\int\frac{p_{\alpha}y+p_{\beta}}{2\left(y^2-1\right)}\frac{y\rd y}{\sqrt{C^2_{5}y^3-B_{1}y^2-B_{2}y+\sigma_{1}}},\label{eq72}\\
\beta(y)&=&\int\frac{p_{\beta}y+p_{\alpha}}{2\left(y^2-1\right)}\frac{y\rd y}{\sqrt{C^2_{5}y^3-B_{1}y^2-B_{2}y+\sigma_{1}}}.\label{eq73}
\end{eqnarray}
Then after the proper integration of the expressions (\ref{eq71})--(\ref{eq73}) and substituting the previously obtained functions $x(t)$ and $y(t)$ into the resulting expressions we obtain the other three solutions $v(t)$, $\alpha(t)$, and $\beta(t)$ of the equations of motion.

Let us note that even in the simplest case of the spherical surface for the internal dynamics we are obtaining that the solutions (\ref{eq70}) and (\ref{eq72})--(\ref{eq73}) are expressed through the special functions (i.e., incomplete elliptic integrals and Jacobi elliptic functions). The detailed analysis of the obtained solutions will be done in the following paper on this subject. 

Nevertheless, we can notice that the translational part of the motion (\ref{eq69}) and (\ref{eq71}) is influenced only by the degrees of freedom connected to the left fictitious gyroscope in our two-polar decomposition, i.e., the coefficients in the corresponding integrals are depending only on the constant of motion $p_{\alpha}$ and are independent of the internal deformation $\lambda$ and the constant of motion $p_{\beta}$. Therefore, in order to study qualitatively the translational motion of our incompressible test body on the spherical surface we can consider the very interesting special case of the purely gyroscopic motion described in the next Section.

\section{Special case of purely gyroscopic motion}

For the particular situation of the purely gyroscopic motion ($\lambda=\beta=0$) we obtain that for the sphere's case using (\ref{eq50a}) the Hamiltonian is given as follows:
\begin{equation}\label{eq74}
\mathcal{H}(q,p)=\frac{\ p^{2}_{u}}{2mR^2}+\frac{\left(p_{v}+\cos u\ p_{\alpha}\right)^2}{2mR^2\sin^2 u}+\frac{p^2_{\alpha}}{2I}+\frac{\varkappa_{1}+2\varkappa_{2}\cos u+\varkappa_{3}\cos^{2} u}{2mR^{2}\sin^2 u}.
\end{equation}
Then the Hamilton's equations of motion (\ref{eq56}) can be written as follows:
\begin{eqnarray}
&&\dot{p}_{u}=-\frac{\partial}{\partial u}\left[\frac{p^2_{v}+\varkappa_{1}+2\cos u\left(p_{v}p_{\alpha}+\varkappa_{2}\right)+\cos^2 u\left(p^{2}_{\alpha}+\varkappa_{3}\right)}{2mR^2\sin^2 u}\right],\label{eq75}\\
&&\dot{u}=\frac{p_{u}}{mR^2},\quad \dot{v}=\frac{p_{v}+\cos u\ p_{\alpha}}{mR^2 \sin^2 u},\quad 
\dot{\alpha}=\frac{\cos u\left(p_{v}+\cos u\ p_{\alpha}\right)}{mR^2 \sin^2 u}+\frac{p_{\alpha}}{I},\qquad\label{eq76}
\end{eqnarray}
whereas $\dot{p}_{v}=\dot{p}_{\alpha}=0$, i.e., $p_{v}=mC_{2}$, $p_{\alpha}=IC_{3}$. Performing the corresponding integration of (\ref{eq75}) we obtain that
\begin{equation}\label{eq77}
p^2_{u}(u)=C^{2}_{1}-\frac{p^2_{v}+\varkappa_{1}+2\cos u\left(p_{v}p_{\alpha}+\varkappa_{2}\right)+\cos^2 u\left(p^{2}_{\alpha}+\varkappa_{3}\right)}{\sin^2 u},
\end{equation}
and therefore, instead of (\ref{eq63}), we have that
\begin{equation}\label{eq78}
E=\frac{C^{2}_{1}}{2mR^2}+\frac{I}{2}C^{2}_{3},
\end{equation}
whereas the parametric solutions of the equations of motion are written as follows:
\begin{eqnarray}
t(x)&=&mR^2\int\frac{\rd x}{\sqrt{A_{1}+2A_{2}x-A_{3}x^2}},\label{eq79}\\
v(x)&=&\int\frac{p_{v}-p_{\alpha}x}{1-x^2}\frac{\rd x}{\sqrt{A_{1}+2A_{2}x-A_{3}x^2}},\label{eq80}\\
\alpha(x)&=&\int\frac{p_{\alpha}x-p_{v}}{1-x^2}\frac{x\rd x}{\sqrt{A_{1}+2A_{2}x-A_{3}x^2}}+C_{3}t(x),\label{eq81}
\end{eqnarray}
where $A_{i}$ are given by the expressions (\ref{eq67}).

\section{Comparison of geodetics and geodesics}

In order to illustrate the general scheme on the example of the geodetic gyroscopic motion on a sphere and compare the obtained solutions with the corresponding geodesics ($\alpha=0$) on the sphere, let us make some additional assumption about the considered system. 

Let us suppose that we have the infinitesimal test body of the unit mass $m=1$ moving on the unit sphere $R=1$ and consider the special physical case when its internal inertia is interrelated with its mass, i.e., $I=mR^2=1$. Apart from that let us also suppose that the translational motion of such an infinitesimal gyroscope on the sphere is performed with the unit speed, i.e., $g_{uu}\dot{u}^2+g_{vv}\dot{v}^2=1$, therefore, the constant $C_{1}=1$. Moreover, let us consider the particular value of the constant $C_2=0$ that is corresponding to the geodesics that are the great circles on the sphere with $v={\rm const}$, i.e., $\dot{v}=0$ and $\dot{u}=1$. 

For the above-described special case of the purely gyroscopic motion on the sphere we have that the constants (\ref{eq67}) are expressed as follows: $A_{1}=1$, $A_{2}=0$, $A_{3}=1+C^{2}_{3}$, where $C^2_{3}$ is describing the internal part of the motion of our infinitesimal gyroscope and can be related to its total energy $E$ by the expression (\ref{eq78}). For illustrative properties let us also suppose that we are considering the situation when $25\%$ of the total energy of the gyroscope are allocated into the translational motion and $75\%$ of the total energy are allocated into the internal motion, i.e., $C_{3}=\pm\sqrt{3}$ and then $E=1/2+3/2=2$, therefore, $A_{3}=4$. 

\begin{figure}[h!]
\centering
\includegraphics[width=8.5cm,keepaspectratio]{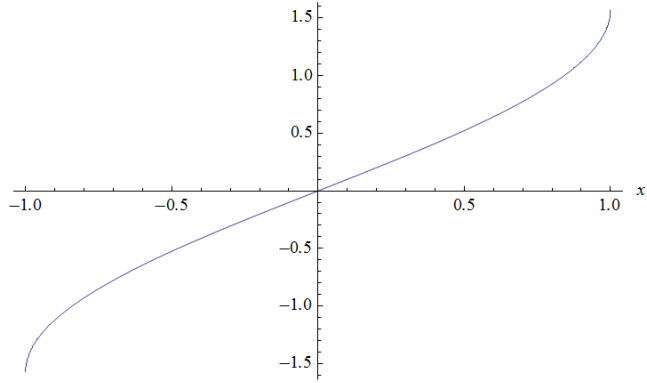}
\caption{The values of function $t(x)$ for geodesics on sphere for $t_{0}=0$: the whole range of variable $x$ is allowed, i.e., $x\in[-1,1]$.}
\end{figure}

Taking into account all the above assumptions, the parametric solutions (\ref{eq79})--(\ref{eq80}) of the geodesic solutions of the equations of motion (when $C_{3}=0$, $A_{3}=1$) are now given as follows:
\begin{equation}\label{eq82a}
t(x)=\int\frac{\rd x}{\sqrt{1-x^2}}=\arcsin\left(x\right)+t_{0},\qquad v(x)=v_{0},
\end{equation}
where $t_{0}$, $v_{0}$ are constants of integration. From the graph of the above function (for $t_{0}=0$) shown on the Figure 1 we see that for the discussed particular kind of geodesics with $v={\rm const}$ (i.e., the meridians that are given as great circles on the sphere) the whole range of the variable $x$ from $x_{\rm min}=-1$ to $x_{\rm max}=1$ (the ultimate values are corresponding to the North and South Poles respectively) is realized during the motion when time is going from $t_{\rm min}=-\pi/2$ to $t_{\rm max}=\pi/2$ (half of the period of the corresponding function $x(t)=\sin t$). Therefore, within the time interval $t\in\left[-\pi/2,\pi/2\right]$ we can invert the function $t(x)$ and obtain that the considered geodesics (meridians) are defined by the following funcitons:
\begin{equation}\label{eq82b}
x(t)=-\cos u(t)=\sin\left(t-t_{0}\right),\qquad v(t)=v_{0},\qquad t\in\left[-\frac{\pi}{2},\frac{\pi}{2}\right].
\end{equation}

\begin{figure}[h!]
\centering
\includegraphics[width=8.5cm,keepaspectratio]{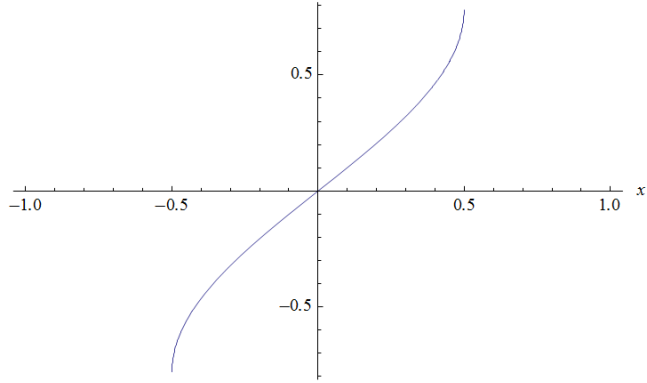}
\caption{The values of function $t(x)$ for geodetics on sphere for $t_{0}=0$: the range of variable $x$ is restricted to $x\in[-0.5,0.5]$.}
\end{figure}

From the other side, the corresponding parametric solution (\ref{eq79}) for the geodetic equations of motion is given in the following form:
\begin{equation}\label{eq82}
t(x)=\int\frac{\rd x}{\sqrt{1-4x^2}}=\frac{1}{2}\arcsin\left(2x\right)+t_{0}.
\end{equation}
From the graph of the solution (for $t_{0}=0$) presented on the Figure 2 we can see that the range of the variable $x$ is now restricted to the interval $[-0.5,0.5]$ which corresponds to the range of the time variable from $t_{\rm min}=-\pi/4$ to $t_{\rm max}=\pi/4$, i.e., again we consider the half of the period of the following solution:
\begin{equation}\label{eq82c}
x(t)=-\cos u(t)=\frac{1}{2}\sin\left[2\left(t-t_{0}\right)\right],\qquad v(t)=v_{0},\qquad t\in\left[-\frac{\pi}{4},\frac{\pi}{4}\right].
\end{equation}

\begin{figure}[h!]
\centering
\includegraphics[width=8.5cm,keepaspectratio]{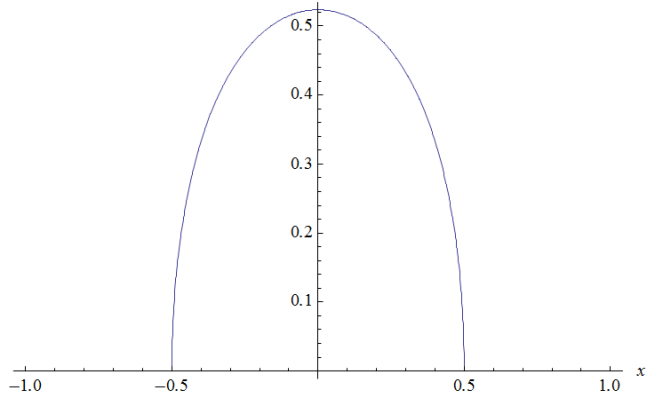}
\caption{The values of function $v_{+}(x)$ for geodetics on sphere for $x\in[-0.5,0.5]$: the geodetic motion is starting and ending on the same meridian with $v_{0}=0$.}
\end{figure}

\begin{figure}[h!]
\centering
\includegraphics[width=8.5cm,keepaspectratio]{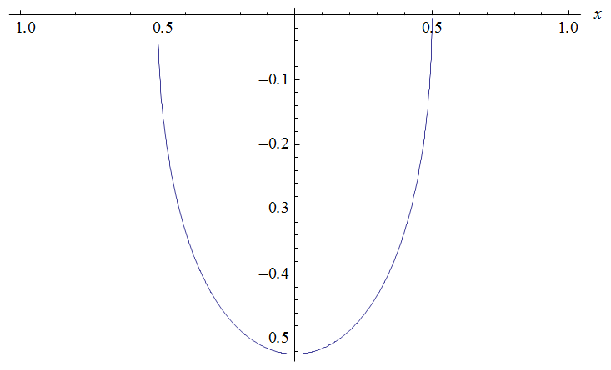}
\caption{The values of function $v_{-}(x)$ for geodetics on sphere for $x\in[-0.5,0.5]$: the geodetic motion is starting and ending on the same meridian with $v_{0}=0$.}
\end{figure}

\begin{figure}[h!]
\centering
\includegraphics[width=8.5cm,keepaspectratio]{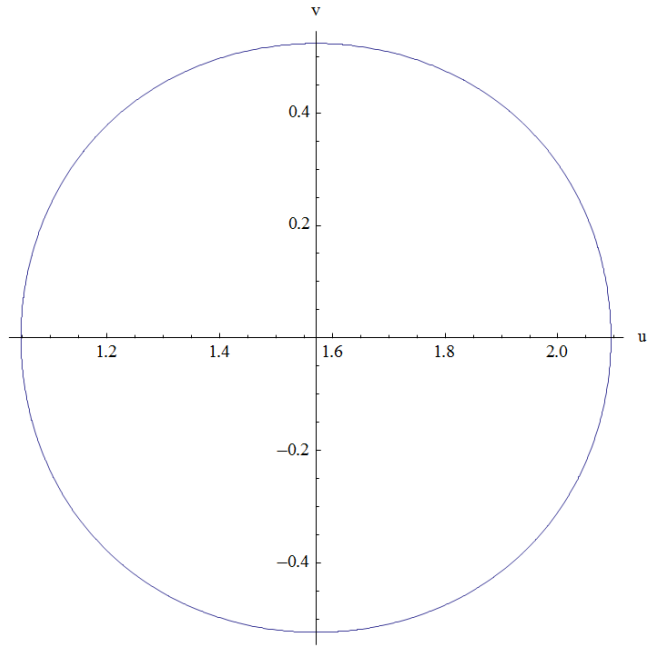}
\caption{In the plane of translational variables $u(t)$ and $v(t)$ the geodetic motion is realized as a small circle on the sphere for $t\in\left[-\pi/4,3\pi/4\right]$.}
\end{figure}

In order to obtain the complete picture of the translational part of the geodetic motion we need to study also the dependency of the other translational variable $v$ on the parameter $x$ which is given as follows (for $C_{3}=\pm\sqrt{3}$):
\begin{equation}\label{eq83}
v_{\pm}(x)=\pm\sqrt{3}\int\frac{x}{x^2-1}\frac{\rd x}{\sqrt{1-4x^2}}=\arctan\left(\frac{\sqrt{1-4x^2}}{\pm\sqrt{3}}\right)+v_{0}.
\end{equation}
From the graph of the obtained solution $v_{+}(x)$ (for $v_{0}=0$) shown on the Figure 3 we can deduce that the geodetic motion is restricted not only in the variable $x$ but also in the variable $v$, i.e., it is realized between the corresponding parallels $x_{\rm min}=-0.5$ ($u_{\rm min}=\pi/3$) and $x_{\rm max}=0.5$ ($u_{\rm max}=2\pi/3$) and corresponding meridians $v_{\rm min}=0$ and $v_{\rm max}=\pi/6$ (half of the period).

In other words, the geodetic motion (for the particular values of the integration constants $t_{0}=0$ and $v_{0}=0$) starts for the time instant $t_{\rm min}=-\pi/4$ on the geodesic (meridian) corresponding to $v_{\rm min}=0$ on the level of the parallel $u_{\rm min}=\pi/3$, then deviates from the original geodesic in the direction of the higher values of the variable $v$ up to the maximal deviation for the time instant $t=0$ when it lands on the meridian $v_{\rm max}=\pi/6$ on the level of the equator ($u=\pi/2$), and then it returns in the symmertical way to the original geodesic ($v_{\rm min}=0$) on which it lands again for the time instant $t_{\rm max}=\pi/4$ on the level of the parallel $u_{\rm max}=2\pi/3$.

The other half of the period is obtained when we consider the solution $v_{-}(x)$ for $C_{3}=-\sqrt{3}$ shown on the Figure 4 that is taken from $x_{\rm max}=0.5$ to $x_{\rm min}=-0.5$ for the time interval $t\in[\pi/4,3\pi/4]$. Glued together both solutions $v_{\pm}(x)$ produce a small circle on the sphere (its graph is shown on the Figure 5, where $\Delta u=2\pi/3-\pi/3=\pi/3$ and $\Delta v=\pi/6+\pi/6=\pi/3$). Therefore, our geodetics are given as planar curves (small circles) on the sphere corresponding to the geodesics that are given as great circles (meridians). 

\begin{figure}[h!]
\centering
\includegraphics[width=8.5cm,keepaspectratio]{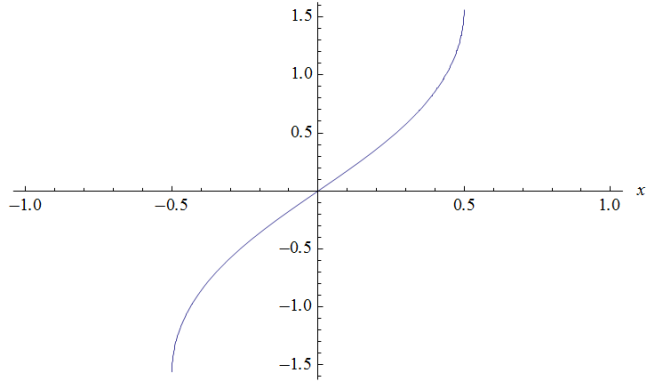}
\caption{The values of function $\alpha_{+}(x)$ for geodetics on sphere for $x\in[-0.5,0.5]$: the infinitesimal gyroscope is rotating all the time in the same direction.}
\end{figure}

\begin{figure}[h!]
\centering
\includegraphics[width=8.5cm,keepaspectratio]{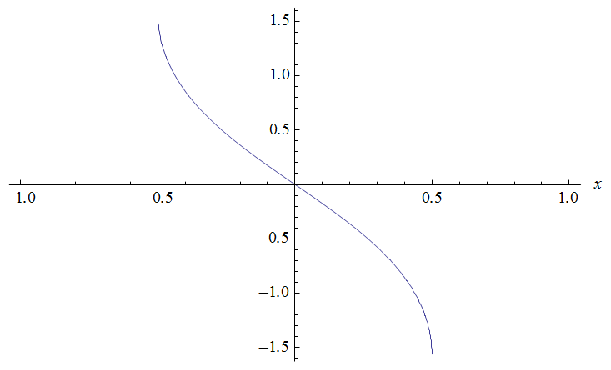}
\caption{The values of function $\alpha_{-}(x)$ for geodetics on sphere for $x\in[-0.5,0.5]$: the infinitesimal gyroscope is rotating all the time in the same direction.}
\end{figure}

And finally, for the internal (rotational) variable $\alpha$ we have that
\begin{eqnarray}
\alpha_{\pm}(x)&=&\pm\sqrt{3}\int\frac{1}{1-x^2}\frac{\rd x}{\sqrt{1-4x^2}}=\alpha_{0}\nonumber\\
&+&\frac{1}{2}\left[\arctan\left(\frac{\left(1+4x\right)}{\pm\sqrt{3\left(1-4x^2\right)}}\right)-\arctan\left(\frac{\left(1-4x\right)}{\pm\sqrt{3\left(1-4x^2\right)}}\right)\right],\qquad\label{eq84}
\end{eqnarray}
where $\alpha_{0}$ is a constant of integration. From the graph of the solution $\alpha_{+}(x)$ (for $\alpha_{0}=0$) presented on the Figure 6 we see that during the geodetic motion our infinitesimal gyroscope is rotating all the time in the same direction taking values from $\alpha_{\rm min}=-\pi/2$ for the time instant $t_{\rm min}=-\pi/4$ up to $\alpha_{\rm max}=\pi/2$ for the time instant $t_{\rm max}=\pi/4$ (half of the period). The other half of the period is obtained when we consider the solution $\alpha_{-}(x)$ for $C_{3}=-\sqrt{3}$ shown on the Figure 7 that is taken from $x_{\rm max}=0.5$ to $x_{\rm min}=-0.5$ for the time interval $t\in[\pi/4,3\pi/4]$. Glued together both solutions $\alpha_{\pm}(x)$ produce one complete revolution in the internal variable $\alpha$ (from $-\pi/2$ to $3\pi/2$) per one complete revolution in the translational variables $u$ and $v$ (shown on the Figure 5). 

Key details of the above discussion on the geodetic motion of the infinitesimal gyroscope on the sphere are summarized in the following table (where in (\ref{eq82c})--(\ref{eq84}) for $t\in[-\pi/4,\pi/4]$ we take the constants $\left(t_{0},v_{0},\alpha_{0}\right)=(0,0,0)$ and $C_{3}=+\sqrt{3}$, whereas for $t\in[\pi/4,3\pi/4]$ we have $\left(t_{0},v_{0},\alpha_{0}\right)=\left(0,0,\pi\right)$ and $C_{3}=-\sqrt{3}$):
\begin{center}
\begin{tabular}{| c | c | c | c | c | c | c | c | c | c |}
\hline
 $t$ & $-\pi/4$ & $-\pi/8$ & $0$ & $\pi/8$ & $\pi/4$ & $3\pi/8$ & $\pi/2$ & $5\pi/8$ & $3\pi/4$\\ 
\hline
 $x$ & $-0.5$ & $-0.35$ & $0$ & $0.35$ & $0.5$ & $0.35$ & $0$ & $-0.35$ & $-0.5$ \\ 
\hline
 $v$ & $0^{\rm o}$ & $22.2^{\rm o}$ & $30^{\rm o}$ & $22.2^{\rm o}$ & $0^{\rm o}$ & $-22.2^{\rm o}$ & $-30^{\rm o}$ & $-22.2^{\rm o}$ & $0^{\rm o}$ \\ 
\hline
$\alpha$ & $-90^{\rm o}$ & $-40.9^{\rm o}$ & $0^{\rm o}$ & $40.9^{\rm o}$ & $90^{\rm o}$ & $139.1^{\rm o}$ & $180^{\rm o}$ & $220.9^{\rm o}$ & $270^{\rm o}$ \\
\hline
\end{tabular}
\end{center}

\section*{Conclusions}\label{rems}

In the present paper we have discussed the mechanics of incompressible test bodies moving in Riemannian spaces with non-trivial curvature tensors. From the very beginning the general scheme has been illustrated with the help of the quite simple but very instructive example of a two-dimensional surface with constant positive Gaussian and mean curvatures, i.e., a sphere, which is embedded into the three-dimensional Euclidean space. Next we have discussed the D'Alembert model of the kinetic energy in the two-polar decomposition of the configuration matrix with some convenient choices of the potential energy. As a result the Hamilton's equations of motion have been formulated and their solutions in the parametric form have been obtained for the incompressible test bodies as well as for the special case of the purely gyroscopic motion. In order to illustrate the obtained solutions the comparison of geodetics and geodesics has been performed and the influence of the internal (gyroscopic) degrees of freedom on the translational ones has been analyzed. We have shown that for geodesics given as great circles (meridians) on the sphere the corresponding geodetics are given as small circles (planar curves) on the sphere (therefore, the motion is restricted in the plane of the translational variables $u$ and $v$). 

As a continuation of the presented work we are planning to analyse more detailly in the following papers the internal part of the motion for which the solutions (\ref{eq70}) and (\ref{eq72})--(\ref{eq73}) are expressed with the help of incomplete elliptic integrals and Jacobi elliptic functions, as well as consider the motion of incompressible test bodies on different and more irregular two-dimensional surfaces embedded into the three-dimensional Euclidean space, e.g., other (apart from spheres) Delaunay and minimal surfaces of constant (including zero) mean curvature (cylinders, catenoids, helicoids, unduloids, nodoids, gyroids, etc.), other (apart from spheres) algebraic surfaces of the second and fourth orders (ellipsoids, pseudo-spheres, tori, etc.), or quite specific but very interesting from the geometrical point of view surface which is called the Mylar balloon \cite{KM_20,Mlad_04,MO_03}.

\section*{Acknowledgements}\label{ackn}

\noindent This paper presents part of the results obtained during the realization of the joint research project \# 32 under the title {\it ``Mechanics of Test Bodies Moving on Curved Membranes''} accomplished within the realm of the scientific cooperation between Bulgarian and Polish Academies of Sciences for the period 2018-2020.

\end{document}